\documentstyle[amssymb,aps,multicol,epsfig,fancyheadings]{revtex}
\begin{document}
\title{Anisotropic conductivity of Nd$_{1.85}$Ce$_{0.15}$CuO$_{4-\delta }$ films at
submillimeter wavelengths}
\author{A. Pimenov$^{1}$, A. V. Pronin$^{1}$, A. Loidl$^{1}$, U. Michelucci$^{2}$,
A. P. Kampf$^{2}$, S. I. Krasnosvobodtsev$^{3}$, V. S. Nozdrin$^{3}$, and D.
Rainer$^{4}$}
\address{$^{1}$Experimentalphysik V, EKM, Universit\"{a}t Augsburg, 86135
Augsburg, Germany\\ $^{2}$Theoretische Physik III, EKM, Universit\"{a}t
Augsburg, 86135 Augsburg, Germany\\ $^{3}$Lebedev Physics Institute,
Russian Acad. Sci., 117942 Moscow, Russia\\ $^{4}$Physikalisches Institut,
Universit\"{a}t Bayreuth, 95440 Bayreuth, Germany}
\maketitle

\begin{abstract}
The anisotropic conductivity of thin Nd$_{1.85}$Ce$_{0.15}$CuO$_{4-\delta }$
films was measured in the frequency range 8 cm$^{-1}<\nu <$ 40 cm$^{-1}$ and
for temperatures 4 K $<T<300$ K. A tilted sample geometry allowed to extract
both, in-plane and c-axis properties. The in-plane quasiparticle scattering
rate remains unchanged as the sample becomes superconducting. The
temperature dependence of the in-plane conductivity is reasonably well
described using the Born limit for a d-wave superconductor. Below $T_{{\rm C}%
}$ the c-axis dielectric constant $\varepsilon _{1c}$ changes sign at the
screened c-axis plasma frequency. The temperature dependence of the c-axis
conductivity closely follows the linear in T behavior within the plane.
\end{abstract}

%\pacs{}

\begin{multicols}{2}

The electron--doped superconductor Nd$_{2-x}$Ce$_{x}$CuO$_{4}$ (NCCO) \cite
{tokura,hidaka} reveals a number of properties, which are rather different
from other cuprate superconductors. It has long been believed that the
superconductivity in this compound is characterized by a s-wave order
parameter (for a review see \cite{fournier}). However, recent experiments,
including phase-sensitive tricrystal\cite{tsuei} and penetration depth
measurements\cite{kokales,prozorov}, strongly support d-wave type symmetry.

In-plane microwave properties of NCCO have been investigated using resonator
techniques \cite{kokales,andreone,anlage,wu}. The infrared conductivity has
been obtained via Kramers-Kronig analysis of reflectivity data\cite
{homes,lupi} and by thin film transmission\cite{choi}. In contrast to a
number of ab-plane experiments, there exists only little information
concerning the c-axis properties of NCCO, which is explained by the
typically small dimensions of the samples along the c-axis. Most experiments
on c-axis dynamics\cite{shibata,heyen} were carried out using
polycrystalline NCCO samples.

Recently we have demonstrated the possibility of using a tilted-sample
geometry to extract the anisotropic conductivity of layered cuprates in the
submillimeter frequency range \cite{tilted}. This method combines the
possibilities of the quasioptical transmission geometry with the high
anisotropy of NCCO which may be estimated by the resistivity ratio $\rho
_{c}/\rho _{ab}\sim 10^{4}$ \cite{o}. In this paper we present the in-plane
and c-axis conductivity of an oxygen reduced Nd$_{1.85}$Ce$_{0.15}$CuO$%
_{4-\delta }$ film (T$_{C}$=16.9K, sample \#A) in the submillimeter
frequency range ($8$ cm$^{-1}<\nu <40\,$cm$^{-1})$ and for temperatures $4$
K $<T<300$ K. Data on an optimally doped film\cite{tilted} (sample \#B, $%
T_{C}=$ 20.1 K) are also discussed.

The films were prepared using a two-beam laser deposition on YSZ substrates%
\cite{preparation}. X-ray analysis showed the c-axis orientation of the
films relative to the crystallographic axes of the substrate. The YSZ
substrate of the presented film (\#A) was tilted from the (001) orientation
by an angle $\alpha =2.6^{o}\pm 0.5^{o}$. Therefore, the film was also
tilted by the same angle from the ideal c-axis orientation. The oxygen
concentration in the deposition chamber was reduced compared to the optimum
value which resulted in a lower $T_{{\rm C}}$ of the film (\#A). The
ac-susceptibility measurements revealed an onset temperature of 16.9 K and a
slightly broader transition width ($\Delta T[10\%-90\%]=2.0$ K) as compared
to the optimally doped film (sample \#B: $\Delta T=0.9$ K \cite{tilted}).

\begin{figure}[tbp]
\centering
\includegraphics[width=8cm,clip]{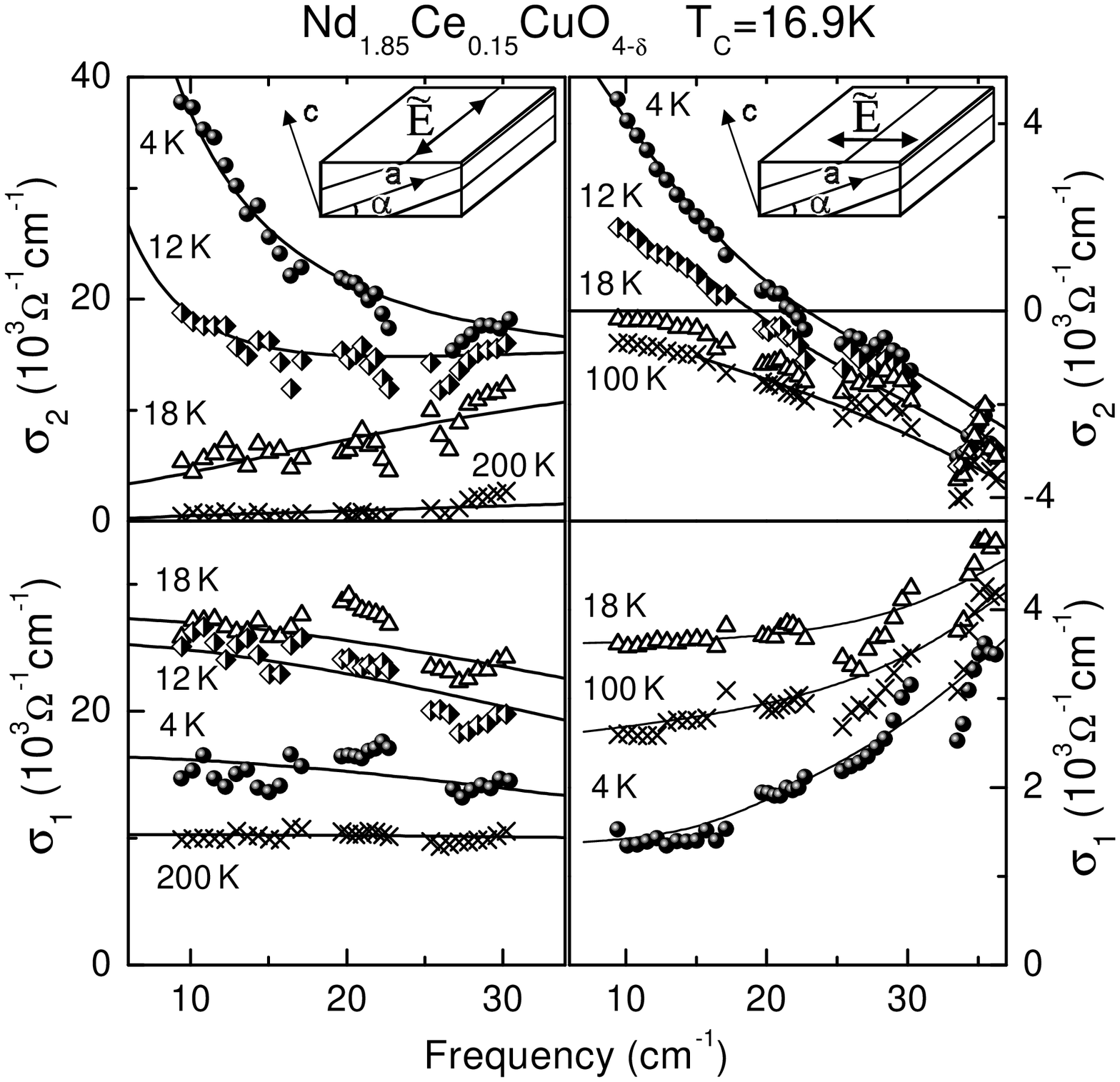}
\vspace*{1mm}
\caption{ Real (bottom panels) and imaginary (top panels) parts of the
complex conductivity of Nd$_{1.85}$Ce$_{0.15}$CuO$_{4-\delta }$ (film \#A)
for different geometries of the transmission experiment as indicated in the
insets. Left panels: ab-plane conductivity. Solid lines are calculated as
explained in the text. Right panels: mixed conductivity. Dashed lines are
guides to the eye. }
\label{sigmix}
\end{figure}

The transmission experiments in the frequency range $6$ cm$^{-1}<\nu <40$ cm$%
^{-1}$ were carried out in a Mach-Zehnder interferometer arrangement \cite
{volkov} which allows both the measurements of transmittance and phase
shift. The properties of the blank substrate were determined in a separate
experiment. Utilizing the Fresnel optical formulas for the complex
transmission coefficient of the substrate-film system, the absolute values
of the complex conductivity $\sigma ^{*}=\sigma _{1}+i\sigma _{2}$ were
determined directly from the observed spectra. Using the tilted sample
geometry at different polarizations of the incident radiation it was
possible to separate the conductivity at a given tilt angle into ab-plane
and c-axis components. The geometry of the experiments is shown in the
insets in Fig.\thinspace \ref{sigmix}.

The conductivity of a tilted sample may be calculated assuming a
free-standing film of thickness $d$ in a uniform electromagnetic field $%
Ee^{-i\omega t}$ parallel to the surface. If the film is thin compared to
the penetration depth, $d\ll \lambda $, then the current and field
distributions may be considered to be uniform. Taking into account the
charges formed at the surface, the following equations can be derived for
the geometry given in the right inset of Fig. \ref{sigmix}:

\begin{equation}
\left\{
\begin{array}{l}
j_{a}=\sigma _{a}[E\cos \alpha -(s/\varepsilon _{0})\sin \alpha ] \\
j_{c}=\sigma _{c}[-E\sin \alpha -(s/\varepsilon _{0})\cos \alpha ]
\end{array}
\right. \quad .  \label{eqmaxw}
\end{equation}

Here $j_{a}$\thinspace $(j_{c})$ is the current density, $\sigma _{a}$%
\thinspace $(\sigma _{c})$ is the complex conductivity in the ab-plane
(along the c-axis), $s$ is the surface charge density, $\varepsilon _{0}$ is
the permittivity of free space, $\omega =2\pi \nu $ is the angular
frequency, and $\alpha $ is the tilt angle. An additional equation, $i\omega
s+j_{a}\sin \alpha +j_{c}\cos \alpha =0$ , follows from the charge
conservation. The effective conductivity of the film can be defined through $%
\sigma _{eff}E=j_{eff}=j_{x}\cos \alpha +j_{y}\sin \alpha $. Solving these
equations one obtains:

\begin{equation}
\sigma _{eff}=\frac{-i\varepsilon _{0}\omega (\sigma _{a}\cos ^{2}\alpha
+\sigma _{c}\sin ^{2}\alpha )+\sigma _{a}\sigma _{c}}{-i\varepsilon
_{0}\omega +\sigma _{a}\sin ^{2}\alpha +\sigma _{c}\cos ^{2}\alpha }\quad .%
\text{ }  \label{eqsig}
\end{equation}

Both $\sigma _{eff}$ and $\sigma _{a}$ can be determined experimentally
using the geometry shown in the right and left insets of Fig. \ref{sigmix},
respectively. Therefore Eq. (\ref{eqsig}) can easily be solved for the
c-axis conductivity. Within the approximation $\alpha \approx \sin \alpha
\ll 1$ and $|\sigma _{a}|\gg |\sigma _{c}|$, Eq. (\ref{eqsig}) may be
simplified to:

\begin{equation}
\sigma _{eff}=\frac{\sigma _{a}(\sigma _{c}-i\varepsilon _{0}\omega )}{%
\sigma _{a}\alpha ^{2}+(\sigma _{c}-i\varepsilon _{0}\omega )}
\label{eqsimpl}
\end{equation}

As discussed previously\cite{tilted}, two excitations may be observed within
the tilted geometry: i) a peak in the real part of the conductivity if $%
Im[\sigma _{a}\alpha ^{2}+(\sigma _{c}-i\varepsilon _{0}\omega _{1})]=0$
which corresponds to the mixed ab-plane/c-axis excitation and ii) the
longitudinal resonance if $Im[\sigma _{c}-i\varepsilon _{0}\omega _{0}]=0$
which corresponds to the c-axis plasma frequency. Both excitations have
been detected for the optimally doped film \#B\cite{tilted}. The c-axis
plasma frequency can be identified in the submillimeter frequency range
also for the reduced Nd$_{1.85}$Ce$_{0.15}$CuO$_{4-\delta }$ film (see
below). Due to the larger tilt angle of the sample \#A the frequency of the
mixed resonance is shifted to $\nu _{1}\sim 200$ cm$^{-1}$ (compared to
$\nu
_{1}\sim 20$ cm$^{-1}$ for sample \#B\cite{tilted}) and occurs as a broad
maximum in the effective conductivity spectra at infrared frequencies \cite
{pronin}.

\begin{figure}[tbp]
\centering
\includegraphics[width=7cm,clip]{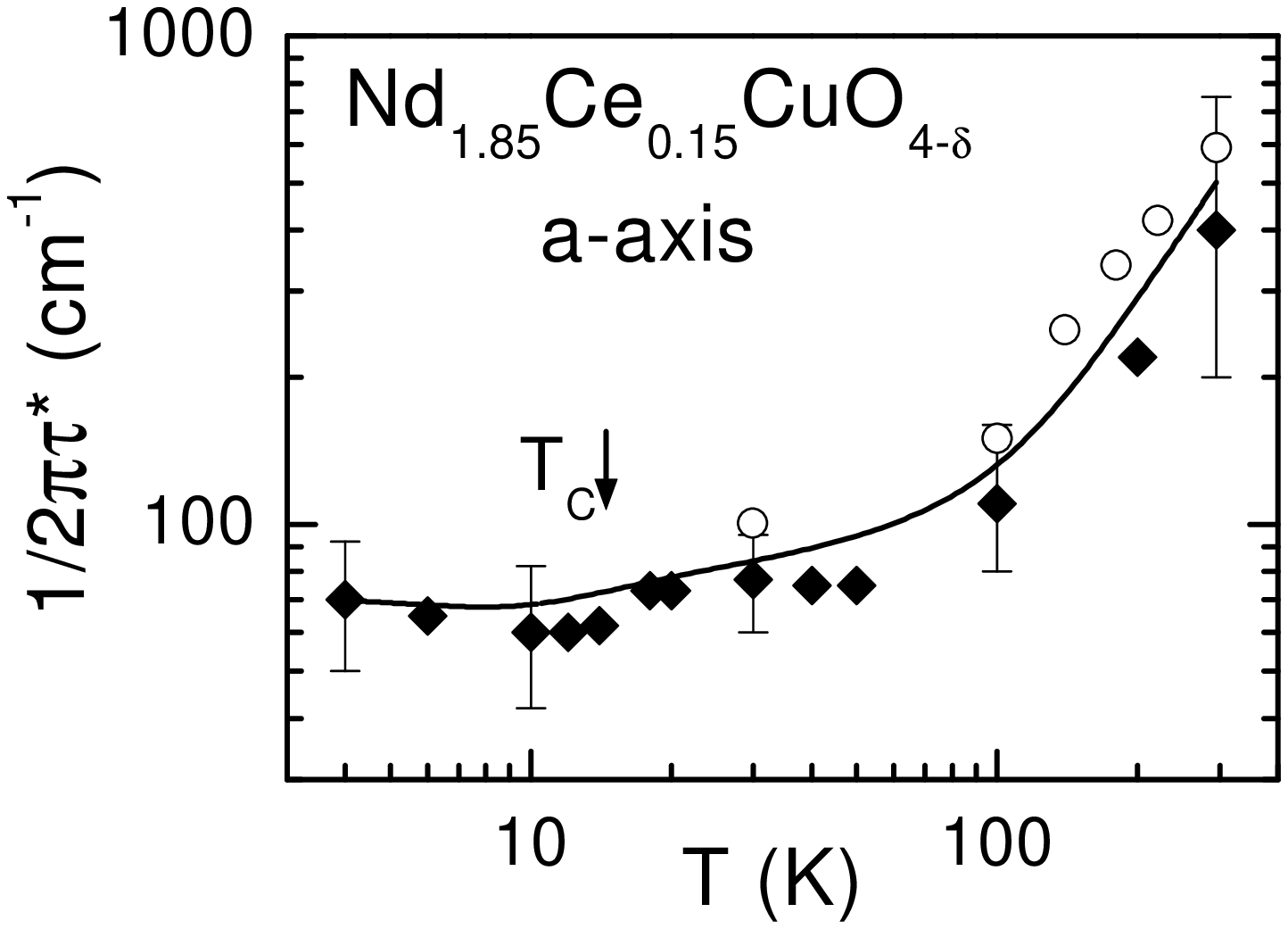}
%\vspace*{1mm}
\caption{ In-plane quasiparticle scattering rate of Nd$_{1.85}$Ce$_{0.15}$CuO%
$_{4-\delta}$ film \#A (full diamonds) as obtained from the Drude analysis
of the complex conductivity data (left panels of Fig.\,\ref{sigmix}). Open
circles: infrared data from Ref.\protect\cite{homes}. Solid line represents $%
1/2 \pi \tau =67$ cm$^{-1} $ for $T<70$ K and $1/2 \pi \tau \propto T$ for $%
T>70$ K. }
\label{figtau}
\end{figure}

Fig.\thinspace \ref{sigmix} shows the complex conductivity of the reduced Nd$%
_{1.85}$Ce$_{0.15}$CuO$_{4-\delta }$ film obtained as described above. The
left panels represent the real (lower frame) and imaginary (upper frame)
parts of the complex conductivity for currents within the CuO$_{2}$ plane.
The real part of the in-plane conductivity $\sigma _{1a}$ is weakly
frequency-dependent in the submillimeter frequency range at all measured
temperatures. This indicates that the quasiparticle scattering rate is
larger than the frequency of the experiment. Consequently, the imaginary
part of the conductivity $\sigma _{2a}$ is nearly zero for high temperatures
but starts to show distinct frequency dependence on approaching the
superconducting transition temperature. From the analysis of $\sigma _{1a}$
and $\sigma _{2a}$ the quasiparticle scattering rate may be estimated using
the Drude expression: $\sigma ^{*}=\sigma _{dc}/(1-i\omega \tau )$. The term
$[i/\omega +\pi \delta (\omega )/2]/[\mu _{0}\lambda _{a}^{2}(T)]$ has to be
added to the Drude expression for $T<T_{{\rm C}}$ in order to account for
the superconducting condensate\cite{ybco}. Here $\lambda _{a}^{2}$ is the
in-plane penetration depth, $\mu _{0}$ the permeability of free space. For
finite frequencies the additional term influences $\sigma _{2a}$ only. The
solid lines in the left panels of Fig.\thinspace \ref{sigmix} were
calculated using the Drude expression extended to temperatures below $T_{%
{\rm C}}$ as described above. The fits allow to estimate the quasiparticle
scattering rate both, below and above $T_{{\rm C}}$. The results are shown
in Fig.\thinspace \ref{figtau} (full diamonds). The scattering rate is
approximately constant for $T\lesssim 70$ K and shows only a small anomaly
(within experimental errors) at $T_{{\rm C}}$. For $T>T_{{\rm C}}$, $1/\tau $
agrees well with the infrared data of Homes {\it et al.}\cite{homes} (open
circles) and has an approximate linear temperature dependence for $T>100$ K.
The absence of the anomalous suppression of the quasiparticle scattering is
in contrast to the results on other cuprate superconductors\cite{bonn}. This
possibly indicates that $1/\tau $ is determined by impurity scattering for $%
T<100$ K. For $T\ll T_{C}$ the $(1/\omega )$ frequency dependence dominates
the imaginary part of the conductivity $\sigma _{2a}$ which allows to
estimate the low-frequency in-plane penetration depth, $\lambda
_{a}(6K)=0.35\mu $. For the optimally doped sample \#B we obtained $\lambda
_{a}(6K)=0.23\mu $ \cite{tilted}.

The right panels of Fig.\thinspace \ref{sigmix} represent the effective
(mixed) conductivity which is described by Eq. (\ref{eqsig}). Using this
equation, the c-axis conductivity may be calculated from the in-plane and
mixed conductivity data. The imaginary part of the mixed conductivity
crosses zero around $\nu \sim 20$ cm$^{-1}$. As will be seen below, this
frequency corresponds to the c-axis plasma resonance.

\begin{figure}[tb]
\centering
\includegraphics[width=7cm,clip]{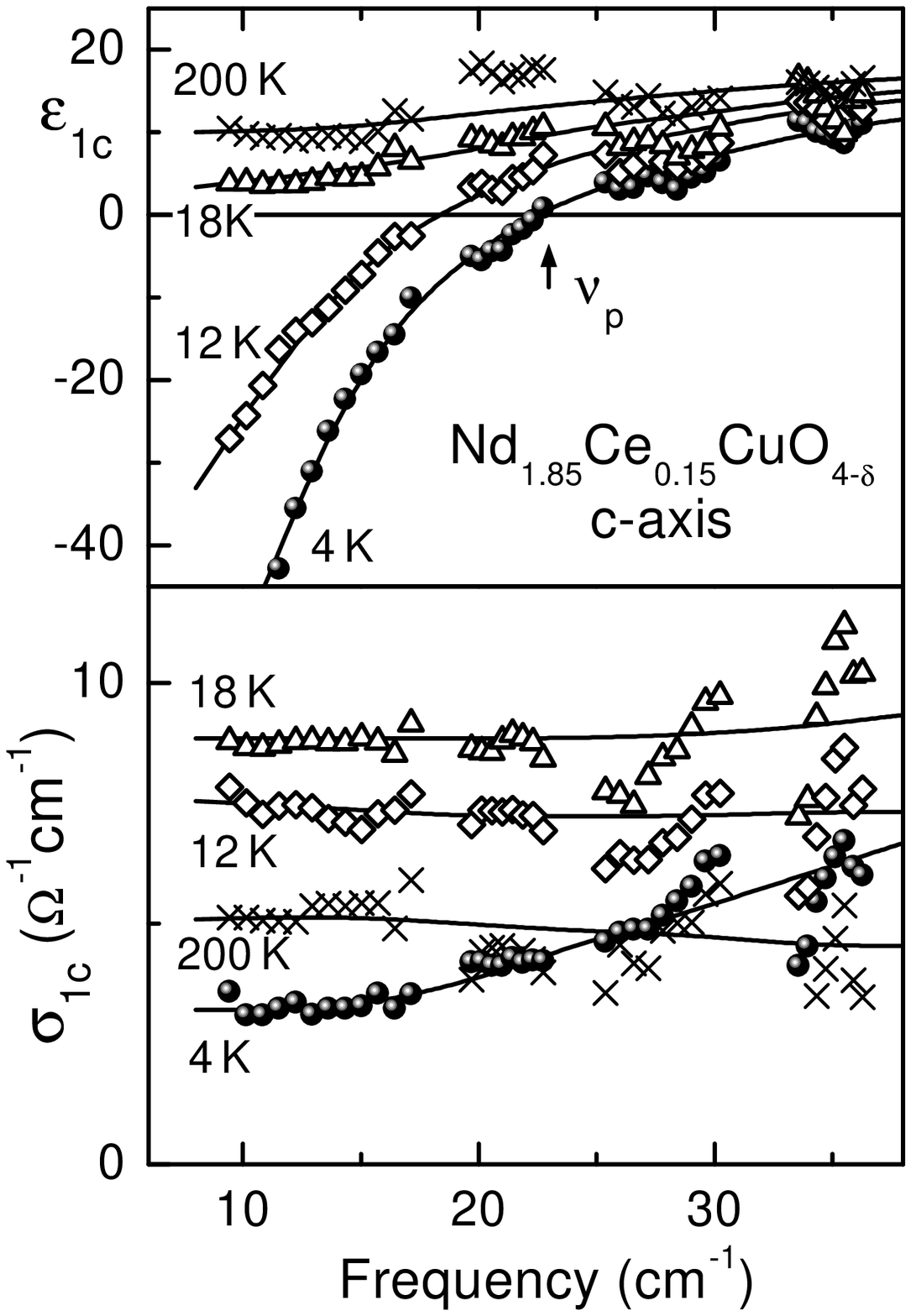}
\vspace*{1mm}
\caption{ Conductivity $\sigma _{1c}$ (bottom panel) and dielectric constant
$\varepsilon_{1c}=-\sigma _{2c}/(\varepsilon _{0}\omega )$ (top panel) of Nd$%
_{1.85}$Ce$_{0.15}$CuO$_{4-\delta}$ (sample \#A) along the c-axis. Lines are
guides to the eye. Arrow indicates the position of the c-axis plasma
frequency.}
\label{caxis}
\end{figure}

Fig.\thinspace \ref{caxis} shows the conductivity $\sigma _{1c}$ and the
dielectric constant $\varepsilon _{1c}=-\sigma _{2c}/(\varepsilon _{0}\omega
)$ of Nd$_{1.85}$Ce$_{0.15}$CuO$_{4-\delta }$ (\#A) along the c-axis. The
real part of the c-axis conductivity (lower frame) is approximately
frequency independent within experimental accuracy and for temperatures well
above $T_{{\rm C}}$. This behavior agrees well with the low-frequency
infrared conductivity of La$_{2-x}$Sr$_{x}$CuO$_{4}$\cite{uchida,kim}, YBa$%
_{2}$Cu$_{3}$O$_{6.7}$\cite{homes93}, and Tl$_{2}$BaCuO$_{6+x}$\cite{basov99}%
. Only at the lowest temperatures $\sigma _{1c}$ does increase with
frequency which most probably reflects the vicinity of the c-axis phonon at $%
\nu \approx 134$ cm$^{-1}$ \cite{heyen,pronin}. The c-axis dielectric
constant is dominated by the high-frequency (phonon) contribution and shows
a weak frequency dependence at high temperatures. As the sample becomes
superconducting, $\varepsilon _{1c}$ reveals a $(1/\omega ^{2})$ behavior,
which gives an estimate of the penetration depth, $\lambda _{c}(6K)=19.2\mu $%
. Consequently, a zero crossing of $\varepsilon _{1c}$ is observed around $%
20 $ cm$^{-1}$ which corresponds to the (screened) plasma frequency $2\pi
\nu _{p}=c/(\lambda _{c}\varepsilon _{\infty }^{1/2})$ where $\varepsilon
_{\infty }\approx 14$ is the high-frequency dielectric constant and $c$ is
the speed of light. For the sample \#B we found $\nu _{p}=12$cm\thinspace $%
^{-1}$ and $\varepsilon _{\infty }\approx 23$ \cite{tilted}.

Assuming Josephson coupling between the CuO$_{2}$ planes, Basov {\it et al.}%
\cite{basov94} suggested a correlation between $\lambda _{c}(0)$ and the
normal-state conductivity $\sigma _{c}(T_{{\rm C}}):\hbar /(\mu _{0}\lambda
_{c}^{2})=\pi \Delta \sigma _{c}(T_{{\rm C}})$. On the basis of this
correlation the results on both NCCO samples give an energy gap $2\Delta
\simeq 30$ cm$^{-1}$. This value may be compared to $2\Delta \simeq 60$ cm$%
^{-1}$ as determined by Raman scattering\cite{hackl}.

\begin{figure}[tb]
\centering
\includegraphics[width=7cm,clip]{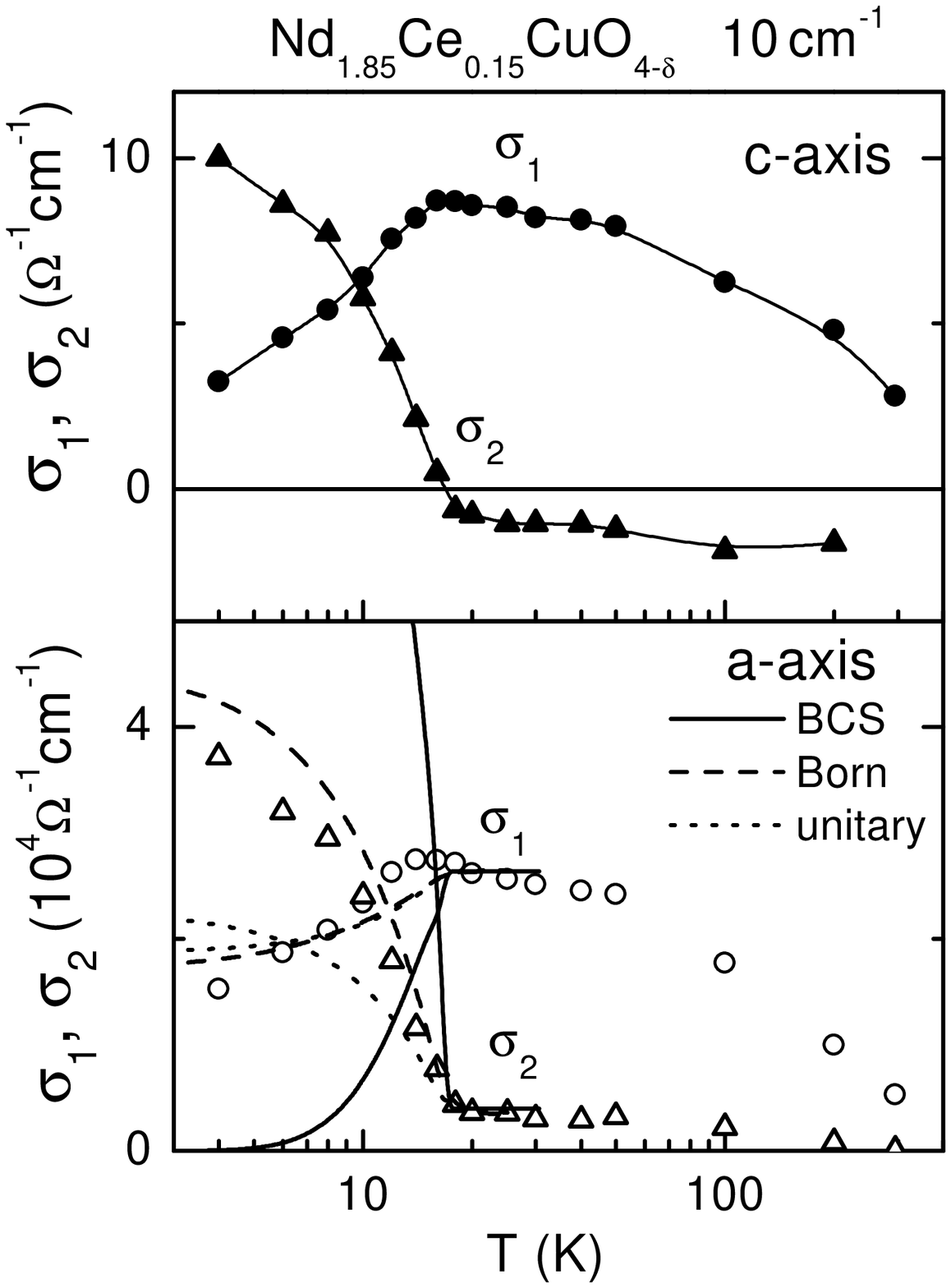}
\vspace*{1mm}
\caption{ Temperature dependence of the complex conductivity of Nd$_{1.85}$Ce%
$_{0.15}$CuO$_{4-\delta}$ film \#A at $\nu =10$ cm$^{-1}$. Upper panel:
c-axis. Lines are guides to the eye. Lower panel: ab-plane. Lines are
calculated according to the s-wave BCS model (solid), Born (dashed) and
unitary (dotted) limits of the d-wave model \protect\cite{graf}. }
\label{tempdep}
\end{figure}

The temperature dependence of the anisotropic conductivity, as measured at $%
\nu =10$ cm$^{-1}$, is represented in Fig.\thinspace \ref{tempdep}. The
lower panel of Fig.\thinspace \ref{tempdep} shows the real and imaginary
parts of the in-plane conductivity. For decreasing temperature, $\sigma
_{1a} $ increases below room temperature, saturates between $T\simeq 100$ K
and T$_{{\rm C}}$ and finally decreases after a slight maximum near $T_{{\rm %
C}}$. A peak near $T_{{\rm C}}$ observed in $\sigma _{1a}$ at microwave
frequencies was recently reported for NCCO by Kokales {\it et al.}\cite
{kokales} and interpreted as possible evidence for suppression of the
quasiparticle scattering. According to Fig.\thinspace \ref{figtau}, our data
suggest a rather temperature independent scattering of quasiparticles below $%
T=100$ K. At high temperatures the imaginary part of the in-plane
conductivity (lower panel of Fig.\thinspace \ref{tempdep}, open triangles)
has values just above the sensitivity limit of the spectrometer. In the
superconducting state $\sigma _{2a}$ abruptly increases reflecting the
formation of the superconducting condensate.

The lower panel of Fig.\thinspace \ref{tempdep} shows the comparison of the
experimental conductivity with theoretical models. As representative
examples we have taken the s-wave BCS expression, as well as Born $(%
\widetilde{\sigma }=0)$ and unitary $(\widetilde{\sigma }=1)$ limits of a
d-wave superconductor\cite{graf}. Here $\widetilde{\sigma }$ is the cross
section of the impurity scattering. All three models calculate a gap value
self-consistently within the weak coupling limit and assume a {\em %
temperature independent} quasiparticle scattering rate of $1/2\pi \tau =65$
cm$^{-1}$. It has to be pointed out, that real and imaginary parts of the
conductivity have to be fitted simultaneously below and above $T_{{\rm C}}$.
This condition leaves {\em no free parameters} within the models. As
documented by the fit results in Fig. \ref{tempdep}, the s-wave curve (solid
line) shows the poorest agreement with the experiment. In contrast, both
limits of the d-wave model describe $\sigma _{1a}$ reasonably well. However,
the unitary limit (dotted) substantially underestimates the imaginary part $%
\sigma _{2a}$. This is probably because in this limit less spectral weight
is shifted to the $\delta $-function at zero frequencies\cite{graf}. The
best description of $\sigma _{2a}$ may be obtained using an intermediate
scattering cross section $\widetilde{\sigma }\simeq 0.2$. Similar results
have been obtained for the optimally doped sample \#B for which $\widetilde{%
\sigma }=0$ gave the best description of the data.

The upper panel of Fig.\thinspace \ref{tempdep} shows the temperature
dependence of the c-axis conductivity of Nd$_{1.85}$Ce$_{0.15}$CuO$%
_{4-\delta }$. Except for the absolute values, these data closely follow the
temperature dependence of the in-plane conductivity. The most prominent
difference is caused by the strong phonon contribution on the c-axis
conductivity, evidenced by a downward shift of $\sigma _{2c}$. Based on the
strong in-plane momentum dependence of the scattering rate and of the
hopping integral, the anisotropic conductivity for high-T$_{{\rm C}}$
cuprates was recently calculated by van der Marel\cite{marel99}, and Xiang
and Hardy\cite{xiang}. Parametrizing the in-plane momentum by an angle $%
\theta $ and using $t_{c}=-t_{\perp }cos^{2}(2\theta )$ for the c-axis
hopping integral, the c-axis conductivity was found to behave as $\sigma
_{1c}\propto T^{3}$ for not too low temperatures\cite{xiang}. The analysis
of Fig.\thinspace \ref{tempdep} shows, that $\sigma _{1c}$ as well as $%
\sigma _{1a}$ for NCCO depend rather linear on temperature below $T_{{\rm C}%
} $. The explanation for this behavior probably is an impurity-induced
angular-independent contribution to $t_{c}$. Following the calculations
described in Refs.\cite{marel99,xiang} this correction does indeed give a
linear temperature dependence of the c-axis conductivity\cite{kampf}.

In conclusion, the anisotropic conductivity of Nd$_{1.85}$Ce$_{0.15}$CuO$%
_{4-\delta }$ films has been obtained using the tilted-sample geometry in
the frequency range $8$ cm$^{-1}<\nu <40$ cm$^{-1}$ and for temperatures 4 K
$<T<300$ K. The in-plane scattering rate is shown to be unchanged as the
sample becomes superconducting. The temperature dependence of the in-plane
conductivity may be reasonably described within the Born limit of a dirty
d-wave superconductor. The c-axis dielectric constant $\varepsilon _{1c}$ is
dominated by a phonon contribution at high temperatures. A zero crossing of $%
\varepsilon _{1c}$ is directly observed below $T_{{\rm C}}$ which
corresponds to the screened c-axis plasma frequency. In contrast to other
cuprate superconductors, the temperature dependence of the c-axis
conductivity closely follows the in-plane behavior.

This work was supported by BMBF (13N6917/0 - EKM) and in part by the
Deutsche Forschungsgemeinschaft through SFB 484.

\end{multicols}


\begin{references}
\bibitem{tokura}  Y. Tokura {\it et al.}, Nature {\bf 337}, 345 (1989).

\bibitem{hidaka}  Y. Hidaka and M. Suzuki, Nature {\bf 338}, 635 (1989).

\bibitem{fournier}  P. Fournier {\it et al.}, in {\it Gap Symmetry and
Fluctuations in High-$T_{{\rm C}}$ Superconductors}, edited by J. Bok {\it %
et al.}, (Plenum Press, NY, 1998), p.145.

\bibitem{tsuei}  C. C. Tsuei and J. R. Kirtley, cond-mat/0002341.

\bibitem{kokales}  J. D. Kokales {\it et al.}, preprint cond-mat/0002300.

\bibitem{prozorov}  R. Prozorov {\it et al.}, preprint cond-mat/0002301.

\bibitem{andreone}  A. Andreone {\it et al.}, Phys. Rev. B {\bf 49}, 6392
(1994).

\bibitem{anlage}  S. M. Anlage {\it et al.}, Phys. Rev. B {\bf 50}, 523
(1994).

\bibitem{wu}  D.-H. Wu {\it et al.}, Phys. Rev. Lett. {\bf 70}, 85 (1993).

\bibitem{homes}  C. C. Homes {\it et al.}, Phys. Rev. B {\bf 56}, 5525
(1997).

\bibitem{lupi}  S. Lupi {\it et al.}, Phys. Rev. Lett. {\bf 83}, 4852 (1999).

\bibitem{choi}  E.-J. Choi {\it et al.}, Phys. Rev. B {\bf 53}, 8859 (1996).

\bibitem{shibata}  H. Shibata and T. Yamada, Phys. Rev. B {\bf 54}, 7500
(1996); ibid. {\bf 56}, 14275 (1997).

\bibitem{heyen}  E. T. Heyen {\it et al.}, Sol. State Comm. {\bf 74}, 1299
(1990).

\bibitem{tilted}  A. Pimenov {\it et al.}, preprint cond-mat/0003404.

\bibitem{o}  Beom-hoan O and J. T. Markert, Phys. Rev. B {\bf 47}, 8373
(1993).

\bibitem{preparation}  V. S. Nozdrin {\it et al.}, Tech. Phys. Lett. {\bf 22}%
, 996 (1996).

\bibitem{volkov}  A. A. Volkov {\it et al.}, Infrared Phys. {\bf 25}, 369
(1985).

\bibitem{pronin}  A. V. Pronin {\it et al.}, unpublished.

\bibitem{ybco}  A. Pimenov {\it et al.}, Phys. Rev. B {\bf 61,} 7039 (2000).

\bibitem{bonn}  D. A. Bonn and W. N. Hardy in {\it Physical Properties of
High Temperature Superconductors V}, edited by D. M. Ginsberg (World
Scientific, Singapore, 1996), p. 7.

\bibitem{uchida}  S. Uchida {\it et al.}, Phys. Rev. B {\bf 53}, 14558
(1996).

\bibitem{kim}  J. H. Kim {\it et al.}, Physica C {\bf 247}, 297 (1995).

\bibitem{homes93}  C. C. Homes {\it et al.}, Phys. Rev. Lett. {\bf 71}, 1645
(1993).

\bibitem{basov99}  D. N. Basov {\it et al.}, Science {\bf 283}, 49 (1999).

\bibitem{basov94}  D. N. Basov {\it et al., }Phys. Rev. {\bf B} 50, 3511
(1994).

\bibitem{hackl}  B. Stadlober {\it et al.}, Phys. Rev. Lett. {\bf 74}, 4911
(1995).

\bibitem{graf}  M. J. Graf {\it et al.}, Phys. Rev. B {\bf 52}, 10588 (1995).

%\bibitem{quinlan}  S. M. Quinlan {\it et al.}, Phys. Rev. B {\bf 53}, 8575
%(1996).

%\bibitem{durst}  A. C. Durst and P. A. Lee, preprint cond-mat/9908182.

\bibitem{marel99}  D. van der Marel, Phys. Rev. B {\bf 60}, 765 (1999).

\bibitem{xiang}  T. Xiang and W. N. Hardy, preprint cond-mat/0001443.

\bibitem{kampf}  U. Michelucci {\it et al.}, unpublished.
\end{references}
\end{document}